\documentclass[pra,onecolumn,aps,nofootinbib,superscriptaddress,notitlepage]{revtex4-1}
\usepackage{latexsym,amssymb,amsfonts,amsmath,amstext,graphicx,bbm,bm,times,hyperref,relsize,dsfont,theorem,times,varioref,enumitem}
\usepackage[dvipsnames]{xcolor}
\usepackage[normalem]{ulem}

\newcommand{\tr}{\mbox{tr}}

\def\tr{\mbox{tr}}
\def\ip#1#2{\langle{#1}|#2\rangle}
\def\bra#1{\langle{#1}|}
\def\ket#1{|{#1}\rangle}

{\catcode`\|=\active
  \gdef\Braket#1{\begingroup
\mathcode`\|32768\let|\BraVert\left<{#1}\right>\endgroup}}
\def\BraVert{\egroup\,\mid\,\bgroup}

\def\lket#1{\vert#1\rangle\hspace{-1mm}\rangle}
\def\lbra#1{\langle\hspace{-1mm}\langle#1\vert}
\def\lbraket#1#2{\langle\hspace{-1mm}\langle#1\vert#2\rangle\hspace{-1mm}\rangle}

\begin{document}
\title{Speeding up Thermalisation via Open Quantum System Variational Optimisation}
\author{Nishchay Suri}
\affiliation{Department of Physics, Indian Institute of Technology Bombay, Mumbai 400076, India}
\affiliation{Department of Physics, Carnegie-Mellon University, Pittsburgh, PA 15213, USA}

\author{Felix C. Binder}
\affiliation{School of Physical and Mathematical Sciences, Nanyang Technological University, 21 Nanyang Link, Singapore 637371, Singapore}

\author{Bhaskaran Muralidharan}
\affiliation{Department of Electrical Engineering, Indian Institute of Technology Bombay, Mumbai 400076, India}

\author{Sai Vinjanampathy}
\email{sai@phy.iitb.ac.in}
\affiliation{Department of Physics, Indian Institute of Technology Bombay, Mumbai 400076, India}
\affiliation{Centre for Quantum Technologies, National University of Singapore, 3 Science Drive 2, Singapore 117543, Singapore}

\date{\today}

\begin{abstract}
Optimizing open quantum system evolution is an important step on the way to achieving quantum computing and quantum thermodynamic tasks. In this article, we approach optimisation via variational principles and derive an open quantum system variational algorithm explicitly for Lindblad evolution in Liouville space. As an example of such control over open system evolution, we control the thermalisation of a qubit attached to a thermal Lindbladian bath with a damping rate $\gamma$. Since thermalisation is an asymptotic process and the variational algorithm we consider is for fixed time, we present a way to discuss the potential speedup of thermalisation that can be expected from such variational algorithms.
\end{abstract}

\maketitle

\section{Introduction}
As quantum technologies are becoming more mature, methods for the exquisite control of quantum systems of macro- and mesoscopic sizes are continuing to gain in importance. Such quantum optimal control aids in solving optimisation problems for closed and open quantum systems, where a given state transformation or unitary gate is implemented within some prescribed constraints. While much has been achieved in this area in recent years there also remains considerable potential for further development~\cite{Werschnik2007,DAlessandro2008,Brif2010,Glaser2015,Koch2016,Borzi2017}.

Here, we contribute an easily accessible description of a family of variational algorithms. Such a numerical approach to quantum control problems is not just useful but also necessary seeing that, in practice, it is rarely possible to solve for optimal control analytically. Effective computational methods are hence indispensable.


There are three general, numerical strategies to obtain optimal driving sequences for quantum evolution. The first strategy involves using standard gradient based techniques, which use steepest descent to optimise the cost function with respect to the control field. An example of gradient based strategies is the so-called GRAPE algorithm~\cite{Khaneja2005}. The second method, exemplified by the CRAB algorithm, involves using non-gradient optimisers (such as Mead-Nelder algorithm~\cite{Avriel2003}) to perform the optimisation~\cite{Caneva2011}. Finally, a variational program may also be set up to optimise the cost function~\cite{Tannor1992}. Here, a cost functional is proposed which depends on the field, among other variational parameters. By asserting that at the extremal points, the first order variation of the cost function vanishes with respect to all variables, one can obtain a set of equations whose solution only has second order deviations in the cost function. This is analogous to the program of deriving Euler-Lagrange equations and will be the object of study in this article. Specifically, we will prove the uniform convergence of a whole class of non-linear algorithms that are based on the variational principle, reviewing previous works~\cite{Tannor1992,Zhu1998,Zhu1998a,Ohtsuki1999,Maday2003,Ohtsuki2004}.

We note two important points about the interplay between variational algorithms and other algorithms: (a) variational algorithms that are seeded by solutions from gradient and non-gradient optimisers outperform variational algorithms seeded by random initial fields \cite{Rach2015}, and (b) variational algorithms have the advantage that they can take large steps at the beginning (where it is expected that the initial condition is far from the optimal field). Conversely, in comparison with gradient based techniques, the same variational algorithms suffer small step sizes as they come close to the optimal solution \cite{Eitan2011}. This slowdown can be mitigated in practice by switching between these two types of algorithms. 

In this article, we review variational optimisation algorithms for closed and open quantum systems, and apply these techniques to the thermalisation of an open quantum system. Such an optimisation program has the potential to inform the design of optimal thermal machines, in particular concerning finite-time, powerful operation. Consider four-stroke quantum Otto engine as an example, with two strokes being unitary (adiabats) and the other two being thermalisation strokes (isochores). While previous works have focused on optimising the adiabats via optimal control (e.g. see~\cite{Stefanatos2014,Bathaee2016,Kosloff2017}) using, for instance, shortcuts to adiabaticity~\cite{DelCampo2014,Zheng2016,Deng2017}, the problem of speeding up thermalisation is equally important since an optimal engine would optimise both the unitary and the thermalisation strokes.

Due to practical considerations, such optimisation has to be performed within experimental constraints. As a consequence of such constraints we typically find ourselves outside the regime of ``good control", where the control Hamiltonian strength can be assumed to be much larger than the strengths of the uncontrolled system operators (i.e., Hamiltonian and Lindbladian). While thermalisation has been considered in the regime of good control before~\cite{Mukherjee2013}, the question of thermalising an open quantum system in the presence of limited resources remains uninvestigated. We rectify this by discussing the optimisation of thermalisation of an open quantum system to a bath with constraints on the controls.

We begin by reviewing variational algorithms for closed quantum systems in section \ref{sec:var_u}. This is followed by an explicit re-derivation of the variational approach to open quantum systems for Gorini-Lindblad-Kossakowski-
Sudarshan (GLKS) evolution in section~\ref{sec:var_open} (see \cite{Ohtsuki2004} for a derivation within the context of two functionals which include this case). A proof of uniform convergence of these algorithms is presented in the appendix. In section~\ref{sec:thermalisation} we then apply these methods to the problem of fast qubit thermalisation.

\section{Variational Control Algorithms for Unitary Evolution}
\label{sec:var_u}
We begin by reviewing the variational approach to desigining optimal pulse sequences for isolated quantum systems. Consider a quantum system, which evolves according to Schr\"{o}dinger's equation ($\hbar=1$, here and throughout; a dot indicates time-derivative)
\begin{align}
i\dot{\ket{\psi}}=H(t)\ket{\psi},\ket{\psi(0)}=\ket{\psi_0},
\end{align}
where the Hamiltonian $H(t)$ is the sum of a time-independent bare Hamiltonian $H_0$ and a time-dependant control $\xi(t)\mu$. This latter part is written without loss of generality as a product of a field $\xi$ and a time-independant operator  $\mu$:
\begin{align}
\label{eq:H}
H(t)=H_0+\xi(t)\mu. 
\end{align}

The primary goal is to drive the quantum system from its initial state $\ket{\psi_0}$ to a desired target state $\ket{\tau}$ in a given fixed amount of time $T$. We assume that the system can reach the the target state $\ket{\tau}$, which amounts to an assumption about the orbit of the algebra generated by $\mu$ and $H_0$ in the initial state $\ket{\psi_0}$ \cite{DAlessandro2008}. Due to the quantum speed limit~\cite{Deffner2017} there is a trade-off between the time $T$ and magnitude and standard deviation of the driving field required to maximise the fidelity $\vert\ip{\psi(T)}{\tau}\vert$ between output and target state. The quantum speed limit for unitary evolution from $\ket{\psi_0}$ to $\ket{\tau}$ is given by~\cite{Mandelstam1945,Margolus1998,Deffner2013}

\begin{align}\label{eq:tqsl}
t\geq t_{\text{QSL}}=\mathcal{L}(\ket{\psi_0},\ket{\tau})\max\left(\frac{1}{E},\frac{1}{\Delta E}\right).
\end{align}
Here, $\mathcal{L}$ is the Fubini-Study distance\footnote{As we extend the analysis to mixed states below, the natural extension of Eq.~\ref{eq:tqsl} in terms of the Bures angle has been shown to not be tight~\cite{Marvian2016,Campaioli2017} and various refinements have been formulated~\cite{Pires2016,Marvian2016,Mondal2016,Mondal2016a,Deffner2017a,Campaioli2017}. However, even for mixed states the intuitive relationship between driving time and energetic resources remains.}, $E=T^{-1}\int_{0}^{T}dt\left(\langle\psi_0\vert H(t)\vert\psi_0\rangle-E_g(t)\right)$ is the average energy, and $\Delta E=T^{-1}\int_{0}^{T}dt\langle\psi_0\vert (H(t)-E(t))^2\vert\psi_0\rangle^{\frac{1}{2}}$ represents the standard deviation of the Hamiltonian averaged over time. $E_g(t)$ is the instantaneous ground state energy of the Hamiltonian and $E(t)$ is the expectation value of Hamiltonian at time $t$.

Now, the question of driving a quantum state from $\ket{\psi_0}$ to $\ket{\tau}$ becomes trivial if the energy is not restricted in some way: If the states are unitarily connected, the driving time can be made arbitrarily short by simply increasing the average energy $E$ and standard deviation $\Delta E$. This simple issue is avoided by placing realistic constraints on the Hamiltonian~\cite{Binder2015a,Binder2016,Campaioli2017,Uzdin2012}. 

In particular, a physically well motivated constraint is a bound on the fluence, defined as $\int_{0}^{T}dt\xi^2(t)$~\cite{Maday2003}. Fluence represents the integrated power that is transmitted from the control pulse to the system and its environment. Thus a secondary goal is to minimise fluence. The two goals -- minimised fluence and target state fidelity -- can be combined into a single cost-function $J$ given by~\cite{Werschnik2007}
\begin{align}\label{J-free}
\begin{split}
J[\ket{\psi},\ket{\chi},\xi(t)] = \langle\psi(T)\vert Q\vert\psi(T)\rangle-2\mathrm{Re}\int_{0}^{T}dt \bra{\chi(t)}\frac{d}{dt}+iH(t)\ket{\psi(t)}\\-\alpha\int_{0}^{T}dt\xi^2(t),
\end{split}
\end{align}
where $Q=\ket{\tau}\bra{\tau}$ is the projector onto the target state. 
In addition to the two above-mentioned goals, the second term enforces the time-dependant Schr\"{o}dinger equation via the introduction of an undetermined, time-dependant Lagrange multiplier $\ket{\chi(t)}$, called the costate. If $\ket{\psi(t)}$ obeys the Schr\"{o}dinger equation at all times, this integral vanishes. While the first term \textit{maximises} state fidelity, the minus sign in the third term enforces \textit{minimisation} of the fluence. The constant $\alpha$ determines the relative importance of the two terms.

The first order variation $\delta J$ is set to zero to obtain a set of equations, namely
\begin{align}
i\ket{\dot{\psi}(t)}=&H(t)\ket{\psi(t)}\text{, with }\ket{\psi(0)}=\ket{\psi_0}\label{eq:uvar1}\\
i\ket{\dot{\chi}(t)}=&H(t)\ket{\chi(t)}\text{, with }\ket{\chi(T)}=Q\ket{\psi(T)}\label{eq:uvar2}\\
\xi(t)=&-\frac{1}{\alpha}\mathrm{Im}\bra{\chi(t)}\mu\ket{\psi(t)}.\label{eq:uvar3}
\end{align}
Note that until now, the discussion is similar to the derivation of standard Euler-Lagrange type equations~\cite{Gelfand1963}. However, Eqns.~\ref{eq:uvar1}-\ref{eq:uvar3} are not directly solvable in the standard way. This is because $\xi(t)$ depends on $\ket{\psi(t)}$ and $\ket{\chi(t)}$, which  in turn depend on the control field $\xi(t)$. Note that the state equation is simply Schr\"{o}dinger's equation with the boundary condition fixed at the origin of time, $t=0$, whereas the costate equation has a boundary condition specified at the final time $T$. This is a generic feature of variational principles~\cite{Gerjuoy1983}.

Several ansatzes have been developed to solve Eqns.~\ref{eq:uvar1}-\ref{eq:uvar3} self-consistently~\cite{Tannor1992,Zhu1998,Zhu1998a,Maday2003}. In particular, the algorithm in reference~\cite{Maday2003} may be stated in terms of the following equations:
\begin{align}
i\ket{\dot{\psi}^{(k)}(t)}=&H^{(k)}(t)\ket{\psi^{(k)}(t)}
\text{, with }\ket{\psi^{(k)}(0)}=\ket{\psi_0}\\
i\ket{\dot{\chi}^{(k)}(t)}=&\tilde{H}^{(k)}(t)\ket{\chi^{(k)}(t)}
\text{, with }\ket{\chi^{(k)}(T)}=Q\ket{\psi^{(k)}(T)}\\
\xi^{(k)}(t)=&(1-\delta)\tilde{\xi}^{(k-1)}(t)-\frac{\delta}{\alpha}\mathrm{Im}\bra{\chi^{(k-1)}(t)}\mu\ket{\psi^{(k)}(t)}\\
\tilde{\xi}^{(k)}(t)=&(1-\eta)\xi^{(k)}(t)-\frac{\eta}{\alpha}\mathrm{Im}\bra{\chi^{(k)}(t)}\mu\ket{\psi^{(k)}(t)} .
\end{align}
Here, $\tilde H:=H_0+\tilde \xi(t)\mu$. Every point $(\delta,\eta)$, determined by the parameters  $0\leq\delta,\eta\leq2$, denotes a unique algorithm\footnote{The algorithms corresponding to references~\cite{Tannor1992}~and~\cite{Zhu1998,Zhu1998a} are located $\delta=1, \eta=0$ and $\delta=\eta=1$, respectively.}. At the beginning the field $\xi^{(1)}(t)$ is initialised randomly. Then, with each iteration $k$, it is updated according to the equations above and the chosen values of the parameters. It has been proven that the cost function $J^{(k)}$ remains positive semidefinite and the algorithm given above converges uniformly, namely $J^{(k+1)}\geq J^{(k)},~\forall~k$~\cite{Zhu1998,Maday2003}.

To summarise, once initial state, target state and form of the total Hamiltonian are chosen, the optimal control field is given by the choice of the set of parameters $(\alpha,\delta,\eta)$. In the next section, we generalise this algorithm to open quantum systems and prove the uniform convergence theorem within that framework. 

\section{Variational Control of Open Quantum Systems}
\label{sec:var_open}
In the previous section, we discussed how several variational algorithms have been designed to maximise the fidelity of a pure state in a target state while simultaneously minimising the fluence, defined as the area under the square of the control field over time. A natural question is whether this analysis can be extended to open quantum systems (OQS)~\cite{Koch2016}. Such extensions have indeed been presented before, extending the algorithm for $\delta=1$ and $\eta=0$ to open systems and non-linear dynamical equations~\cite{Reich2012} and to open systems at the same point in parameter space~\cite{Bartana1997}. In reference~\cite{Ohtsuki2004}, the authors considered a set of non-Hermitian Hamiltonians and presented a general OQS Krotov algorithm alongside the proof of uniform convergence. In this section, we discuss the OQS generalisation of the whole parameter family of variational algorithms described in the previous section (i.e., $0\leq\delta,\eta\leq2$) explicitly for the GLKS equation (see below) and prove their uniform convergence.

We begin by describing the problem for OQS. As before, the primary goal is to evolve an initial state $\rho_0$ to a target state $\tau$ during a fixed time $T$, employing a control Hamiltonian identical to the previous example, namely $H(t)=H_0+\xi(t)\mu'$. The evolution is given by the standard Markovian master equation for the system interacting with a large Markovian environment -- the GLKS equation
\begin{align}\label{eq:Lindblad}
i\dot{\rho}=[H(t),\rho(t)]+i\sum_{k}\left(L_k\rho(t) L^{\dag}_k -\frac{1}{2}\{L^{\dag}_kL_k,\rho(t)\}\right).
\end{align}
Here, the quantum state is given by the density matrix $\rho(t)$. $\{\circ,\circ\}$ is the standard anti-commutator, and $L_k$ are Lindblad operators. This equation is completely positive and trace preserving, and is a standard model for quantum systems embedded in Markovian baths~ \cite{Breuer2002}. We hence make the assumption that the GLKS equation is a sufficient model for the dynamics we wish to describe. 

We begin by moving to Liouville space, which entails writing the density matrix $\rho(t)$ as a vector $\lket{\psi}$ composed by stacking the columns of $\rho(t)$~\cite{Mukamel1999}. This means that the various terms in the GLKS equation transform according to the formula $B\rho C\rightarrow C^{*}\otimes B\lket{\psi}$. Using this and representing the initial state as $\rho_0\rightarrow \lket{\psi_0}=\lket{\psi(0)}$, we can write the GLKS equation as what is known as the Liouville equation
\begin{align}
\label{eq:Liouville}
i\lket{\dot{\psi}}&= A(t)\lket{\psi}\text{, with}\\
\label{eq:A}
A(t)&=I\otimes H(t)-H^{*}(t)\otimes I+i\sum_{k}\left( L^{T}_{k}\otimes L_k -\frac{1}{2}\left[I\otimes L^{\dag}_kL_k+L^T_kL^*_k\otimes I \right]\right).
\end{align}
As evident in this second equation, the matrix $A(t)$ consists of Hermitian part $X(t)$ and an anti-Hermitian part $iY$. All of the control is in $X(t)$. It is easily seen that if the evolution of the quantum system was unitary, then $Y=0$ and the evolution equation is just Schr{\"o}dinger's equation in a higher dimensional space. In that case, the corresponding variational control theory is identical to the analysis in the preceding section as verified below. On the other hand, a non-zero $Y$ corresponds to a non-Hermitian Hamiltonian. This represents the contractive semigroup structure of the GLKS evolution.

Analogous to Eq.~\ref{J-free}, we propose a cost function 
\begin{align}\begin{split}\label{J-open}
J[\lket{\psi},\lket{\chi},\xi(t)] = \vert\lbraket{\psi(T)}{\tau}\vert^2-2\mathrm{Re}\int_{0}^{T}dt \lbra{\chi(t)}\frac{d}{dt}+iA(t)\lket{\psi(t)}\\-\alpha\int_{0}^{T}dt\xi^2(t).
\end{split}
\end{align}
Note here, that the standard Hermitian inner product between two Liouville states is given by $\lbraket{\rho}{\sigma}\equiv\tr(\rho^{\dag}\sigma)$. With this,
$J$ is understood exactly as before and similarily to the unitary case, setting the first order variation $\delta J=0$ produces three equations:
\begin{align}
i\lket{\dot{\psi}(t)}=&A(t)\lket{\psi(t)}
\text{, with }\ket{\psi(0)}=\lket{\psi_0},\\
i\lket{\dot{\chi}(t)}=&A^{\dag}(t)\lket{\chi(t)}
\text{, with }\ket{\chi(T)}=Q\lket{\psi(T)},\\
\xi(t)=&-\frac{1}{\alpha}\mathrm{Im}\lbra{\chi(t)}\mu\lket{\psi(t)},
\end{align}
where $Q=\lket{\tau}\lbra{\tau}$ and $\mu$ represents the Liouville superoperator corresponding to the control Hamiltonian $\xi(t)\mu'\rightarrow\xi(t)\mu$. Like before, we have to algorithmise to solve these equations. In a similar spirit to the preceding section, we propose the following algorithm
\begin{align}
\label{lkrotov1}
i\lket{\dot{\psi}^{(k)}(t)}=&A^{(k)}(t)\lket{\psi^{(k)}(t)}
\text{, with }\lket{\psi^{(k)}(0)}=\lket{\psi_0}\\
\label{lkrotov2}
i\lket{\dot{\chi}^{(k)}(t)}=&\tilde{A}^{(k)\dag}(t)\lket{\chi^{(k)}(t)}
\text{, with }\lket{\chi^{(k)}(T)}=Q\lket{\psi^{(k)}(T)}\\
\label{lkrotov3}
\xi^{(k)}(t)=&(1-\delta)\tilde{\xi}^{(k-1)}(t)+\frac{\delta}{\alpha}\mathrm{Im}\lbra{\chi^{(k-1)}(t)}\mu\lket{\psi^{(k)}(t)}\\
\label{lkrotov4}
\tilde{\xi}^{(k)}(t)=&(1-\eta)\xi^{(k)}(t)+\frac{\eta}{\alpha}\mathrm{Im}\lbra{\chi^{(k)}(t)}\mu\lket{\psi^{(k)}(t)}. 
\end{align}
Here, the cost function can be indexed with the iteration as $J^{(k)}\equiv J[\lket{\psi^{(k)}},\lket{\chi^{(k)}},\xi^{(k)}(t)]$, just like in the unitary case.  $A^{(k)}(t)$ and $\tilde A^{(k)}(t)$ depend on the control fields $\xi^{(k)}(t)$ and $\tilde \xi^{(k)}(t)$, respectively, according to Eqns.~\ref{eq:H},\ref{eq:Liouville},\ref{eq:A}. As before the initial field $\xi^{(1)}(t)$ is selected at random. The full algorithm is illustrated in Fig.~\ref{fig:algorithm}. Now, we are ready to state following theorem whose proof is given in Appendix~\ref{appendix:proof1}.

\emph{Theorem 1.}~(\cite{Ohtsuki2004}) The algorithm given in Eqns.~\ref{lkrotov1}-\ref{lkrotov4} is uniformly convergent -- i.e., $J^{(k+1)}\geq J^{(k)}~\forall\;k\geq1,0\leq\delta,\eta\leq 2$.

Note that the parameters $[\delta,\eta]$ can also be varied with each step so that the algorithm achieves convergence in the smallest number of steps (see \cite{Ohtsuki2004} for details).

\begin{figure}[ht]
\resizebox{0.7\columnwidth}{!}{%
\includegraphics{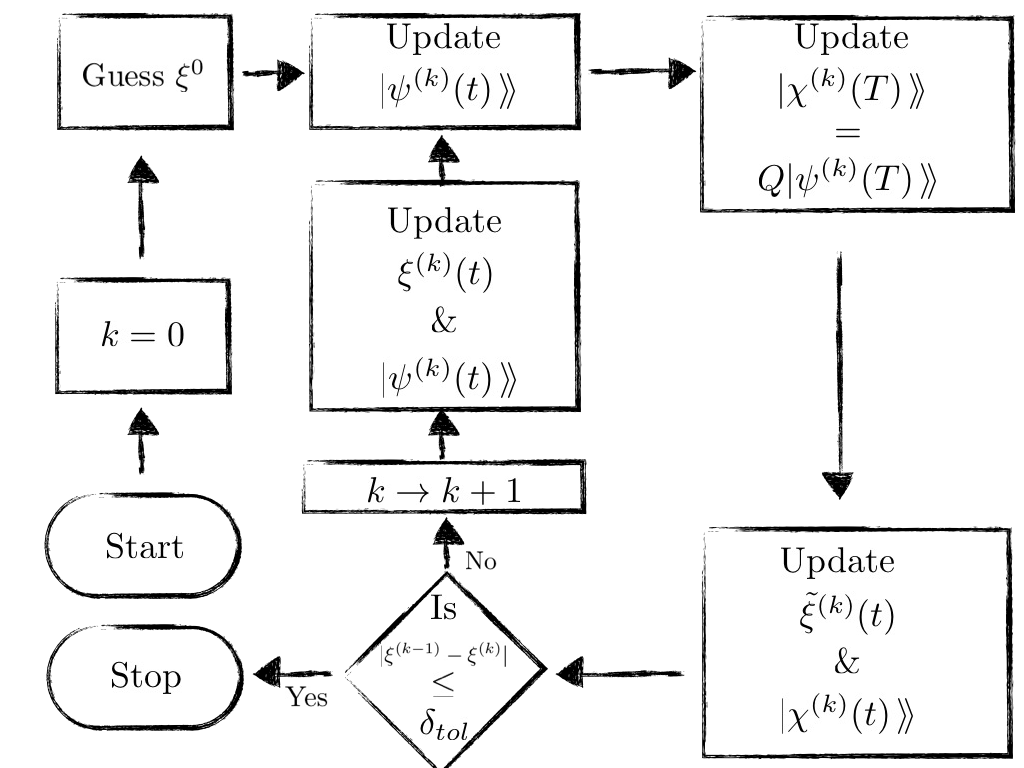}}
\caption{The steps of the open quantum system variational algorithm that is detailed in Eqns.~\ref{lkrotov1}-\ref{lkrotov4} is depicted in the figure. After an initial field $\xi^0$ is guessed, all subsequent steps proceed according to the steps outlined. An error tolerance $\delta_{tol}$ is specified to terminate the algorithm. Alternatively, a $k_{max}$ can be specified for finite number of runs after replacing the decision in the flowchart with $k\geq k_{max}$.} 
\label{fig:algorithm} 

\end{figure}

\section{Application: Speeding up the Thermalisation of a Single Qubit}
\label{sec:thermalisation}
To make use of the algorithm derived in the previous section we now consider thermalisation dynamics of a system attached to a bath -- a paradigmatic example of high relevance to both non-equilibrium thermodynamics and quantum optimal control. For an out-of-equilibrium quantum system which is placed in contact with a heat bath the time required for full thermalisation is usually infinite. The time it takes to reach a neighbourhood of the Gibbs state, on the other hand, is finite. To be concrete, we define the $\varepsilon$-neighbourhood of state $\tau$ to be all those states $\sigma$ for which a suitable distance $D(\sigma,\tau)$ does not exceed $\varepsilon$. As in Eq.~\ref{J-free}, we use the trace distance for $D$ because it easily yields an analytical expression for the present purposes (Eq.~\ref{eq:teps})\footnote{Other distance measures may equally be used. In particular, the relative entropy, despite not being a proper distance measure, carries appeal as an alternative candidate due to its relationship to the non-equilibrium free energy.}.

\begin{figure}[ht]
\resizebox{0.65\columnwidth}{!}{%
\includegraphics{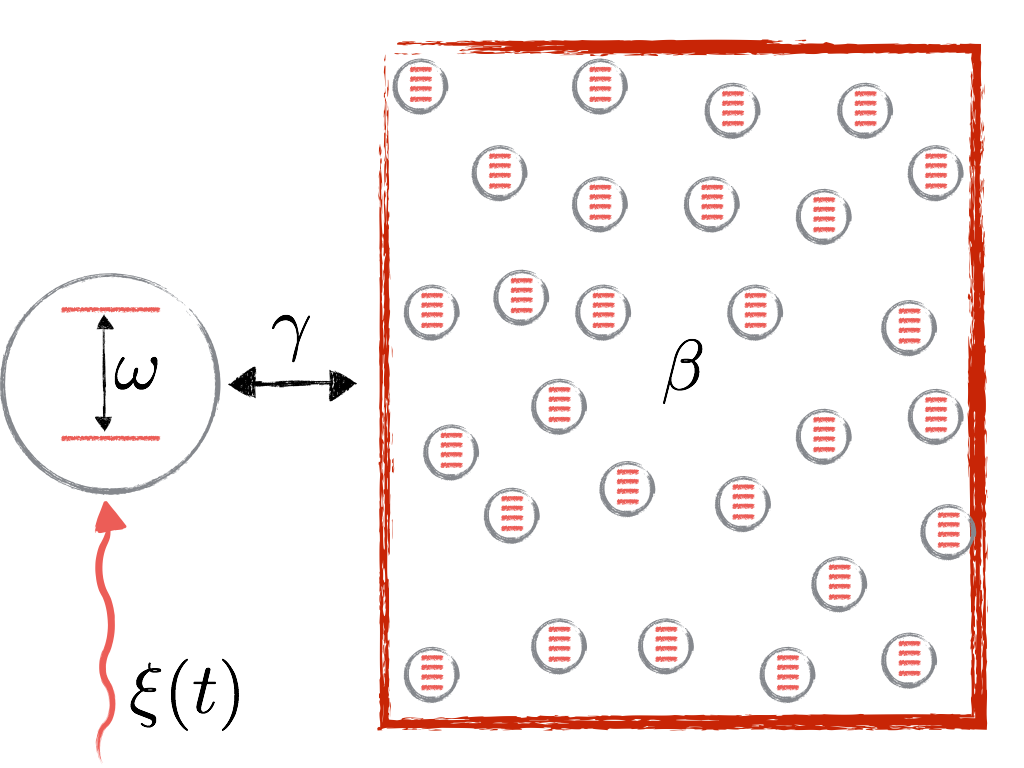}}
\caption{A schematic of speeding up thermalisation of a quantum system. A qubit is in contact with a heat bath at temperature $\beta^{-1}$. The dynamics of this qubit is described by a Markovian master equation with rate $\gamma$ in addition to being subjected to a Hamiltonian $H_0+\xi(t)\mu'$. This field $\xi(t)$ which is optimised such that the thermalisation is faster than a thermalisation timescale (calculated below) which is proportional to $\gamma^{-1}$.}
\label{fig:schematic}
\end{figure}

The time to reach the $\varepsilon$-neighbourhood under free evolution from the initial state, called the $\varepsilon$-free time and denoted by $T^{\varepsilon}_{\text{free}}$, is estimated below. For a given $T^{\varepsilon}_{\text{free}}$, an optimal control field that speeds up thermalisation while minimising fluence may be sought. We will apply the open quantum system variational algorithm above and demonstrate optimal pulse sequences that speed up thermalisation for one qubit. The setup is represented in Fig.~\ref{fig:schematic} for the simple example of a qubit.

We consider a single qubit whose energy gap is $\omega$ connected to a bath at inverse temperature $\beta$. In addition to being in contact with a thermal bath, the qubit is subject to control via a control Hamiltonian $\xi(t)\sigma_x$. Its net evolution is given by
\begin{align}
\displaystyle\frac{d\rho(t)}{d t}=-i\left[\frac{\omega}{2}\sigma_z+\xi(t)\sigma_x,\rho(t)\right]+ \mathcal{D}[\sqrt{\gamma l_+}\sigma_{+}] \rho(t)+\mathcal{D}[\sqrt{\gamma l_-}\sigma_{-}] \rho(t),
\label{lindyeq}
\end{align}
 where $\mathcal{D}[P]\rho:=P\rho P^{\dagger}-\frac{1}{2}\{P^{\dag}P,\rho\}$ represents the standard GLKS evolution, and $l_{+}=\frac{1}{e^{\omega\beta}-1}$ and $l_{-}=\frac{e^{\omega\beta}}{e^{\omega\beta}-1}$ represent the contact of the qubit with a Bosonic bath. $T^{\varepsilon}_{\text{free}}$ can be estimated to be
 \begin{align}
 \displaystyle T^{\varepsilon}_{\text free}=-\frac{1}{2\gamma}\log\left(\frac{-\alpha+\sqrt{\alpha^2+16\varepsilon^2\beta^2}}{2\beta^2}\right),
 \label{eq:teps}
 \end{align}
 where  $\alpha= r^2_x(0)+r^2_y(0)$ and $\beta=r_z(0)+r_{fp}$ in terms of the Bloch vector $\vec{r}(0)$ of $\rho_0$; $(0,0,-r_{fp})$ corresponds to the thermal state. Notice that $\displaystyle\lim_{\varepsilon\rightarrow0}T^{\varepsilon}_{\text free}\rightarrow\infty$, as expected. Further details are provided in Appendix~\ref{appendix:time}.
 
With this estimate for how long it takes a quantum system to evolve to the $\varepsilon$-neighborhood of the thermal state, we can seek quantum control solutions to speed up the thermalisation by a fixed factor (since Krotov is a fixed time algorithm) using our algorithm. To this end, we seek control solutions that speed up the thermalisation by taking an initial state to a state inside the $\varepsilon$-ball in a time $T=T^{\varepsilon}_{\text free}/s$. Here $s$ represents the speedup of the thermalisation. Note that the control algorithm will not produce a viable solution if an initially thermally hot state is in contact with a cold bath (such that both Gibbs states share common eigenvectors), in agreement with other work~\cite{Mukherjee2013}. Likewise, constraints on the field will also fail to produce good algorithms if too little time is given (i.e., if $s$ is too big). To understand this, note that any implicit constraint on the Hamiltonian strength (unspecified by the algorithm) is a constraint on $t_{\text QSL}$, and hence the algorithm will fail in that case, too~\cite{Caneva2009}. In other cases, the open systems variational algorithm produces an optimal field that minimises fluence and maximises $\vert\tr(\rho(T)\tau)\vert^2$. 
 
\begin{figure}[ht]
\resizebox{0.45\columnwidth}{!}{%
\includegraphics{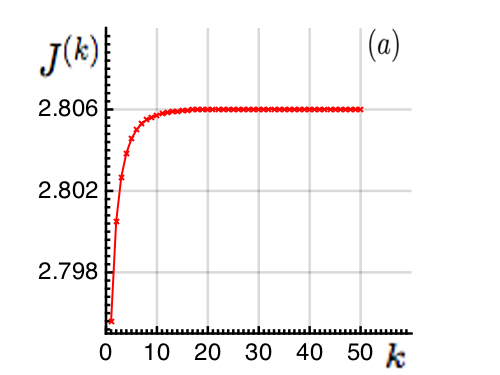}}
\resizebox{0.45\columnwidth }{!}{%
\includegraphics{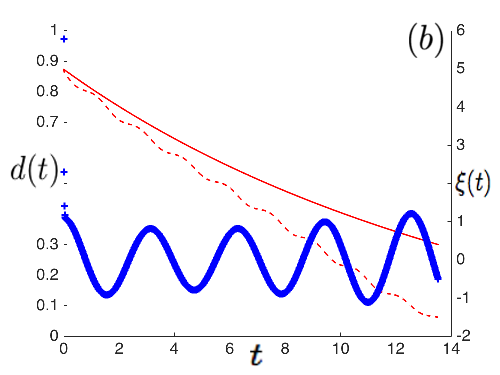}}\\
\caption{An example of the open quantum system Krotov algorithm is presented. In figure (a), the cost function $J^{(k)}$ is plotted against the iteration index $k$, showing the uniform convergence of the algorithm. In figure (b), the trace distance without control (red, solid line) and with control (red, dashed line) is presented on the left vertical axis against time. The field is plotted (plus sign) on the right vertical axis. The details of the optimisation are presented in the text.}
 \label{fig:plots}

\end{figure}

In Fig.~\ref{fig:plots}, we illustrate an example of the variational algorithm employed to speed up thermalisation by a constant factor. We note that unless the Hamiltonian, and hence the control field $\xi(t)$ are constrained, the quantum speed limit is meaningless. Instead of constraining the Hamiltonian, we here simply fix the time $T$ to be smaller than the $T^{\varepsilon}_{\text free}$ by fixing $s>1$, we seek optimal solutions that speed up thermalisation by the given factor $s$. We considered a random initial state, given by 
 \begin{align}
 \rho(0)=\left( \begin{array}{cc}
0.5 & 0.19i \\
-0.19i & 0.5 \end{array} \right),
 \end{align}
 and wish to thermalise it to a bath whose target (thermal) state is given by 
  \begin{align}
\tau=\left( \begin{array}{cc}
0.4 & 0 \\
0 & 0.6 \end{array} \right).
 \end{align}
Focussing on the dynamics specified by Eq.~\ref{lindyeq}, we chose the damping rate associated with the generalised amplitude damping map to be $\gamma=0.1$, which means $T^{\epsilon}_{free}=27.0573$ for a choice of  $\epsilon=0.1$ (both in natural units). We chose $\omega=2$ and $s=2$, meaning we were seeking a halving of the thermalisation time, and ran the open system algorithm one hundred times with $\eta=1.5$ and $\delta=1.5$. The parameter $\alpha$, which sets the relative strength of the fidelity to fluence, was set to $10^{-3}$. The results are presented in Fig.~\ref{fig:plots}. As a simple verification of the theorem, it can be seen that the evolution is uniform convergent. Furthermore, the thermalisation time is improved significantly by the application of the optimal control field. The optimal field is also plotted as a function of time and the fluence minimisation exhibits itself in the field being of small amplitude at all times. 

We emphasise the difference between this variational approach to controlling thermalisation and the approach in reference~\cite{Mukherjee2013}. There, the authors considered two scenarios. In the first one, they assumed that they were in the regime of good control, which means that the effective Hamiltonian strength can be taken to be infinite. This is the reason for the analysis to only invoke the dissipative part of the dynamics where the time taken to go from the initial state to the final state only depended on the Lindbladian dissipators. In the second scenario, the authors considered control Hamiltonians with finite strengths. In this case, the total Hamiltonian was expressed as $H(t)=H_0(t)+mH_c(t)$, where $H_c$ is the control Hamiltonian and $m$ represents its strength. The CRAB algorithm was then used to optimise $H_c(t)$ for various values of $m$. The central difference of this approach with the present one is that while the CRAB technique places a roof on the strength of the Hamiltonian, it does not penalise fluence so long as the field at any instance of time is smaller than that roof. Hence the total fluence in the optimal solutions in the aforementioned work is a result of a fundamentally different optimisation than the one presented here.
 
\section{Summary and Discussion}
In this article we have reviewed a family of variational algorithms for the time-optimal control of open quantum systems and proved their convergence. This extends previous works for optimal control of pure states under unitary evolution~\cite{Tannor1992,Zhu1998,Zhu1998a,Maday2003}, reviews their extension to open systems dynamics~\cite{Ohtsuki2004}, and complements other approaches to the optimal control of open quantum systems~\cite{Bartana1993,Bartana1997,Schulte-Herbrueggen2011,Floether2012,Reich2012,Goerz2014,Koch2016}.

It should be noted that while we prove convergence for these variational algorithms we have not addressed the question of controllability. A system is called controllable if all possible target states can be reached with the available control terms. In the unitary case, controllability is determined by the control terms in the Hamiltonian alone (Eq.~\ref{eq:H})~\cite{Koch2016,Borzi2017}. In the open system case, however, it is clear that the situation is more intricate~\cite{Dive2015,Koch2016}. For instance, basic intuition is correct in asserting that driving towards a pure state (i.e., cooling) cannot be accomplished if the uncontrolled open systems dynamics have a strong tendency to mix the system, as in the case of unital maps. In such cases, no (weak) Hamiltonian controls can achieve controllability (If the control is strong compared to the open systems dynamics the system may of course be driven close to a pure state on short time scales).


We demonstrated the potential of the algorithm for the example of fast thermalisation, complementing alternative methods to achieve the same goal~\cite{Cavina2017,Mukherjee2013}. This offers tantalising projects for further use of numerical optimisation in the area of quantum thermodynamics~\cite{Kosloff2013,Goold2016,Vinjanampathy2016,Millen2016}, not least for the optimisation of engine cycles.

\section*{Acknowledgements}
The authors thank Ronnie Kosloff for helpful comments. FCB acknowledges support by the National Research Foundation of Singapore (Fellowship  NRF-NRFF2016-02).  BM acknowledge support from IIT Bombay SEED grant and ISRORESPOND grant. SV is acknowledges support from an IITB-IRCC grant number 16IRCCSG019 and by the National Research Foundation, Prime Minister's Office, Singapore under its Competitive Research Programme (CRP Award No. NRF-CRP14-2014- 02).

%

\clearpage
\appendix

\section{Proof of Theorem 1}
\label{appendix:proof1}
Consider the difference $ \Delta J^{( k)}\equiv J^{(k+1)}-J^{(k)}$. Assuming that the evolution at each point obeys the GLKS equation, $\Delta J^{(k)}$ is given by 
\begin{align}\label{deltaJ}
\begin{split}
\Delta J^{( k)}=& \bigg\{ \lbra{\psi^{(k+1)}(T)} Q \lket{\psi^{(k+1)}(T)} - \alpha \int_{0}^{T} dt [\xi^{(k+1)}(t)]^2  \bigg\}  \\
&-\bigg\{ \lbra{\psi^{(k)}(T)}Q\lket{\psi^{(k)}(T)}- \alpha \int_{0}^{T} dt [\xi^{(k)}(t)]^2 \bigg\}.
\end{split}
\end{align}
Next, using Eqs.(\ref{lkrotov1},\ref{lkrotov2},\ref{lkrotov3},\ref{lkrotov4}), the inner product $\lbra{\psi^{k+1}(T)-\psi^k(T)}Q\lket{\psi^{k+1}(T)-\psi^k(T)}$ may be expressed as follows
\begin{align}\label{dif}
\begin{split}
&\lbra{\psi^{k+1}(T)}Q\lket{\psi^{k+1}(T)} - \lbra{\psi^{k}(T)}Q\lket{\psi^{k}(T)}=\\
 &\;\;\lbra{\psi^{k+1}(T)-\psi^k(T)}Q\lket{\psi^{k+1}(T)-\psi^k(T)}  +2 Re \lbra{\psi^{k+1}(T) - \psi^{k}(T)}Q\lket{\psi^{k}(T)}.
 \end{split}
\end{align}
Substituting for the difference $\Delta J^{( k)}$  in Eq.~\ref{deltaJ} with the RHS of Eq.~\ref{dif}, we obtain
\begin{align}
\begin{split}
\Delta J^{( k)}=&\lbra{\psi^{k+1}(T)-\psi^k(T)}Q\lket{\psi^{k+1}(T)-\psi^k(T)} +2 Re \lbra{\psi^{k+1}(T) \\&- \psi^{k}(T)}Q\lket{\psi^{k}(T)}- 
\alpha \int_{0}^{T} dt( [\xi^{(k+1)}(t)]^2- [\xi^{(k)}(t)]^2).
\end{split}
\end{align}
Now, using the fact that $2 Re \lbra{\psi^{k+1}(T) - \psi^{k}(T)}Q\lket{\psi^{k}(T)}=2 Re \lbraket{\psi^{k+1}(T) - \psi^{k}(T)}{\chi^{k}(T)}$ and the fundamental theorem of calculus, we write
\begin{align}
\begin{split}
2 Re \lbraket{\psi^{k+1}(T) - \psi^{k}(T)}{\chi^{k}(T)}=&2 Re \int_{0}^{T}dt~\lbraket{\frac{d\psi^{(k+1)}(t)-d\psi^{(k)}(t)}{dt}}{\chi^{(k)}(t)}\\&+\lbraket{\psi^{(k+1)}(t)-\psi^{(k)}(t)}{\frac{d\chi^{(k)}(t)}{dt}}.
\end{split}
\end{align}
Then, substituting the Liouville equation for the evolution of the state and costate, we can write
\begin{align}
2 Re \lbraket{\psi^{k+1}(T) - \psi^{k}(T)}{\chi^{k}(T)}=2 Re\int_{0}^{T}dt~i\{\lbra{\psi^{(k+1)}(t)}(\tilde{A}^{(k+1)\dag}-A^{(k)\dag})\lket{\chi^{(k)}(t)}-\nonumber\\\lbra{\psi^{(k)}(t)}(A^{(k)\dag}-\tilde{A}^{(k)\dag})\lket{\chi^{(k)}(t)}\}
\end{align}
After some simplification, we can write $\Delta J^{( k)}$ as
\begin{align}
\begin{split}
\Delta J^{( k)}= & \lbra{\psi^{k+1}(T) -\psi^k(T)}Q\lket{\psi^{k+1}(T)-\psi^k(T) }\\ &+\alpha \int_0^T dt \bigg( \frac{2}{\delta} - 1 \bigg) (\xi^{k+1} - \tilde{\xi}^k)^2 + \bigg( \frac{2}{\eta} - 1 \bigg) (\tilde{\xi}^{k} - {\xi}^k)^2.
\end{split}
\end{align}
The RHS of the equation above is clearly positive semidefinite for $0\leq\delta,\eta\leq2$. Hence, for open system evolution given by Eq.~\ref{eq:Lindblad}, the algorithms given by Eqns.~\ref{lkrotov1}-\ref{lkrotov4} are uniformly convergent, i.e., $J^{(k+1)}\geq J^{(k)}~\forall k\geq 1,0\leq\delta,\eta\leq2$.  QED.

\section{Calculating the $\varepsilon$-free time for single qubit thermalisation}
\label{appendix:time}

\begin{figure}[ht]
\begin{center}
\resizebox{0.5\columnwidth}{!}{%
\includegraphics{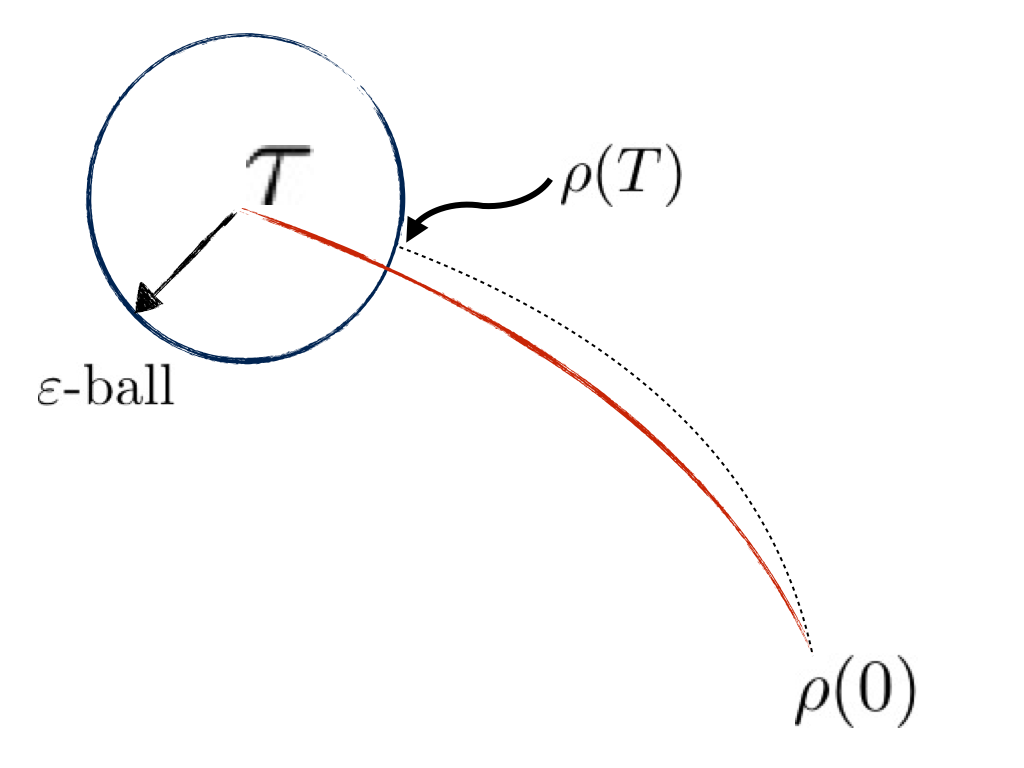}}
\caption{A visualisation of the $\varepsilon$-ball around the target state and the optimisation of the algorithm. Though it is desirable that the initial state $\rho(0)$ be taken to the target state $\tau$, often owing to constraints no such solution is available. As a alternative, we can demand that the quantum state be evolved with an optimal pulse to the $\varepsilon$-neighbourhood of the target state. The ideal trajectory is given in red (solid) line and the optimal realistic trajectory is given by the black (dotted) line.}
 \label{EBall}
\end{center} 
\end{figure}

This section provides a derivation of Eq.(\ref{eq:teps}). The $\varepsilon$-free time is defined as the minimum time it takes to reach the $\varepsilon$-ball of a given state in absence of external control. This is illustrated in Fig. (\ref{EBall}).

Now to proceed with the calculation of this time, we consider the Lindblad equation
\begin{align}
\frac{d\rho(t)}{dt}=-i[H(t),\rho(t)] + \sum_k L_{k}\rho(t)L^{\dag}_{k}-\frac{1}{2}\{L^{\dag}_kL_k,\rho(t)\}.
\end{align}
Following reference~\cite{Mukherjee2013}, we can always write the Lindblad operators as traceless operators $L_k=\gamma^{\frac{1}{2}}_{k}~\mathbf{l}_k\cdot\mathbf{\sigma}$ and the Lindblad equation as
\begin{align}
\dot{\mathbf{r}}=2\mathbf{h}\times \mathbf{r} +2\sum_k\gamma_k\Bigg( \text{Re}[(\mathbf{l}_k.\mathbf{r})\mathbf{l}^{*}_k]-\mathbf{r}+i(\mathbf{l}_k\times\mathbf{l}^{*}_k)\Bigg).
\end{align}
For a generalised amplitude damping map (thermal map) this becomes
\begin{align}
\dot{\mathbf{r}}=(0,0,\omega)^{T}\times\mathbf{r} -\gamma_1(r_x(t),r_y(t),0)^{T} -\gamma_2(0,0,r_z(t))^{T}-\gamma_3(0,0,1)^{T},
\end{align}
where $\gamma_1=\frac{\gamma}{2r_{fp}}$, $\gamma_2=2\gamma_1$ and $\gamma_3=\gamma$. The solution to this set of differential equations is

\begin{align}
r_x(t)=e^{-\gamma_1 t}[r_x(0)\cos(\omega t)-r_y(0)\sin(\omega t)]\\
r_y(t)=e^{-\gamma_1 t}[r_y(0)\cos(\omega t)+r_x(0)\sin(\omega t)]\\
r_z(t)=-\frac{\gamma_3}{\gamma_2}+e^{-\gamma_2 t}\left(r_z(0)+\frac{\gamma_3}{\gamma_2}\right)\Rightarrow\\
r_z(t)=-r_{fp}+e^{-\gamma_2 t}\left(r_z(0)+r_{fp}\right)
\end{align}

Now, we also note that if there are two density matrices $\rho=\frac{1}{2}(I+\mathbf{r}.\sigma)$ and $\varsigma=\frac{1}{2}(I+\mathbf{s}.\sigma)$, then the trace distance $D_1$ is given by 
\begin{align}
D_1(\rho,\varsigma):=\frac{1}{2}\Vert\mathbf{r}-\mathbf{s}\Vert=\frac{1}{2}[(r_1-s_1)^2+(r_2-s_2)^2+(r_3-s_3)^2]^{\frac{1}{2}}.
\end{align}
The state of the qubit at time $t$ is given by $\mathbf{r}(t)=(r_x(t),r_y(t),r_z(t))^T$ with the elements of the vector defined above. We wish to estimate the trace distance from the target thermal state, which is written in terms of the Bloch vector as $\mathbf{r_{*}}=(0,0,-r_{fp})^{T}$ with $r_{fp}=\frac{e^{\omega\beta}-1}{e^{\omega\beta}+1}$ where $\omega$ is the energy gap of the qubit.
 We can compute the trace distance between $\mathbf{r}$ and $\mathbf{r_*}$ at any time
\begin{align}
D_1(\rho,\rho_{*}):=\Vert\mathbf{r}(t)-\mathbf{r_*}\Vert= \frac{e^{-\gamma_1t}}{2}\left( r^2_x(0)+r^2_y(0)+e^{-2\gamma_1t}(r_z(0)+r_{fp})^2\right)^{\frac{1}{2}}.
\end{align}

Now, consider the trace distance $D_1(\rho,\rho_{*}):=\Vert\mathbf{r}(t)-\mathbf{r_*}\Vert$. We can contemplate the $\varepsilon$-free time, defined as the minimum time needed to get to within distance $\epsilon$ of the fixed point:

\begin{align}
T^\varepsilon_{free}:=\operatornamewithlimits{argmin}_{t}
(\Vert\mathbf{r}(t)-\mathbf{r_*}\Vert\leq\varepsilon)
\end{align}
It can be computed from the analytic formula above to be Eq. (\ref{eq:teps}).

\end{document}